\documentclass[twocolumn,tighten]{aastex6}
\usepackage{hyperref,url}
\usepackage{natbib,graphicx,amsmath}

\shorttitle{Variable stars in NGC 6652}
\shortauthors{Salinas et al.}

\begin{document}

\title{New variable stars in NGC 6652 and its background Sagittarius stream$^{\dagger}$}

\author{Ricardo Salinas\altaffilmark{1}, A. Katherina Vivas\altaffilmark{2} and Rodrigo Contreras Ramos\altaffilmark{3,4}}
          
\altaffiltext{1}{Gemini Observatory, Casilla 603, La Serena, Chile; rsalinas@gemini.edu}
\altaffiltext{2}{Cerro Tololo Interamerican Observatory, National Optical Astronomy Observatory, Casilla 603, La Serena, Chile}
\altaffiltext{3}{Millennium Institute of Astrophysics, Av. Vicu\~na Mackenna 4860, 782-0436 Macul, Santiago, Chile}
\altaffiltext{4}{Instituto de Astrof{\'i}sica, Pontificia Universidad Cat\'olica de Chile, Av. Vicu\~na Mackenna 4860, 782-0436 Macul, Chile}          

\begin{abstract}

We conducted a variable star search on the metal-rich Galactic globular cluster NGC 6652 using archival Gemini-S/GMOS data. We report the discovery of nine new variable stars in the NGC 6652 field, of which we classify six as eclipsing binaries and one as  SX Phoenicis stars, leaving two variables without classification. Using proper motions from \textit{Gaia} DR2 and \textit{HST}, albeit with some uncertainties, we find that the cluster, the field, and the background Sagittarius stream, have 3 of these variables each. We also reassess the membership of known variables based on the \textit{Gaia} proper motions, confirming the existence of one RR Lyrae star in the cluster.

\end{abstract}

   
\section{Introduction}\label{sec:intro}

\let\thefootnote\relax \footnotetext{$^{\mathrm{\dagger}}$Based on observations obtained at the Gemini Observatory, which is operated by the Association of Universities for Research in Astronomy, Inc., under a cooperative agreement with the NSF on behalf of the Gemini partnership: the National Science Foundation (United States), the National Research Council (Canada), CONICYT (Chile), Ministerio de Ciencia, Tecnolog\'{i}a e Innovaci\'{o}n Productiva (Argentina), and Minist\'{e}rio da Ci\^{e}ncia, Tecnologia e Inova\c{c}\~{a}o (Brazil).}

The study of variable stars is paramount to our understanding of the distance scale, and the stellar structure and evolution itself \citep[e.g.][]{catelan15}. The variable star content of star clusters is particularly relevant since it allow us to study ensembles of these stars at essentially the same distance, helping to understand the systematic errors that affect the different stellar distance indicators.

Stellar variability in Galactic globular clusters (GCs) is one of the oldest branches of astronomy \citep{pickering89,bailey02} and by the mid 90s it was thought that the census of their variable star content was almost complete \citep{suntzeff91}. This situation changed spectacularly with the introduction of image subtraction techniques \citep[e.g][]{alard98}, which resulted in hundreds of new variable stars found even in previously well-studied clusters \citep[e.g.][]{kaluzny04,contreras05}. Even though stellar variability has now been studied using image subtraction techniques in many GCs \citep[e.g.][]{salinas05,catelan06,arellano17}, there are still GCs which have not been studied in this way, or where improved observational techniques have pushed even deeper into the cores of GCs \citep{skottfelt13,salinas16}. In this work we focus on one of those understudied clusters, NGC 6652, with the unusual treat for Galactic variability studies of using data from a 8m-class telescope.

NGC 6652 (C\,1832-330, RA=18 35 45.6, Dec=-32 59 26.6), is a fairly metal rich, ([Fe/H]=--0.96 and [Fe/H]=--0.85 in the \citealt{zinn84} and \citealt{carretta97} scales, respectively), old Galactic GC \citep[11.7 Gyr,][]{chaboyer00}, associated either with the inner halo \citep{chaboyer00} or the outer parts of the bulge \citep{rossi15}. It has heliocentric distance of 10 kpc, and a Galactocentric distance of 2.7 kpc \citep[][2010 edition]{harris96}. It has a core radius of 0.1\arcmin\, and a tidal radius of 6.3\arcmin\, \citep{trager93}. This cluster has been a frequent subject of study mainly because of the high number of X-ray sources it harbors \citep{predehl91,heinke01,coomber11,engel12,decesar15}. Optical variability studies have been conducted by \citet{hazen89} using photographic plates, and therefore more sensitive to discovery in the outer parts of the cluster, and recently by \citet{skottfelt15}, using EMCCD data restricted to the very inner 45\arcsec$\times$45\arcsec\, of the cluster, leaving a significant portion of the cluster still unexplored with modern techniques. Also both studies are relatively shallow, leaving unstudied the variability below the main sequence turnoff (MSTO).

\begin{figure*}[t]
\centering
\vspace{-1.5ex}
\begin{minipage}{.47\textwidth}
\includegraphics[width=0.99\textwidth]{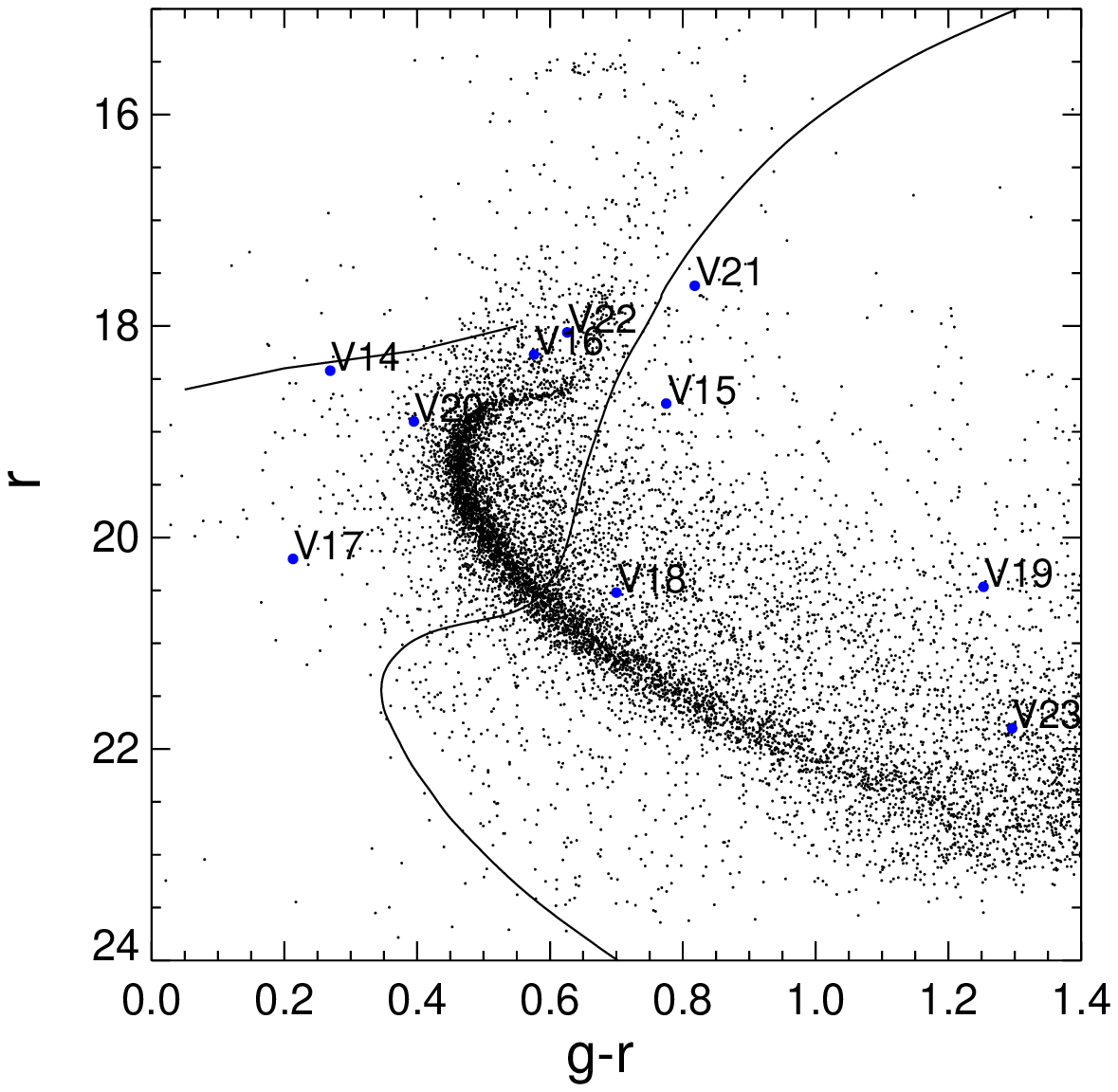}
\end{minipage}
\begin{minipage}{.47\textwidth}
\includegraphics[width=0.99\textwidth]{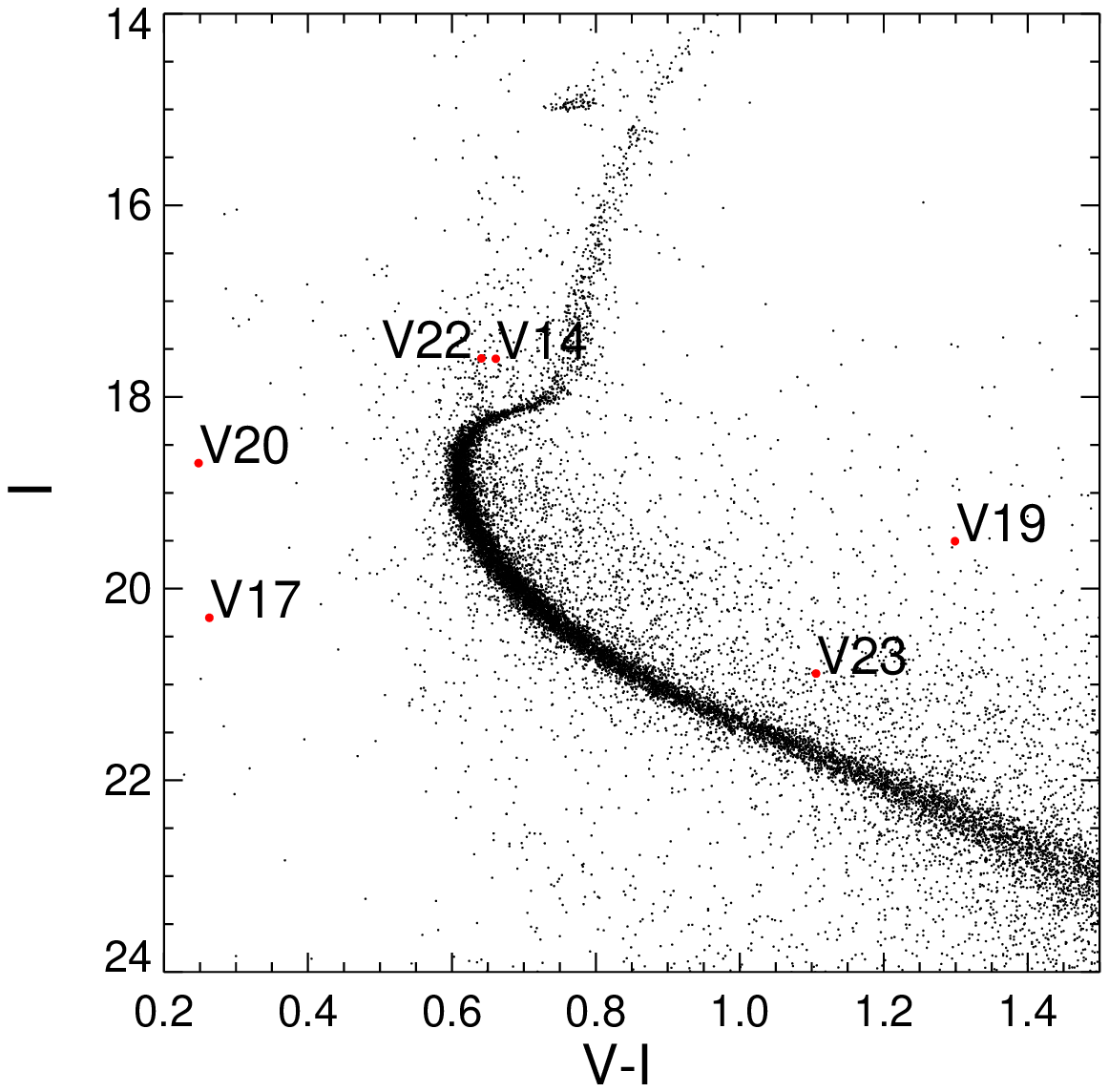}
\end{minipage}
\caption{{\bf Left panel:} GMOS photometry of NGC 6652. Intensity weighted magnitude and colors for the discovered variables V15 to V23 together with known variable V14 are shown in blue symbols. An isochrone from the \citet{dotter08} models with 11 Gyr and [Fe/H]=$-1.3$, consistent with the background Sgr population and shifted to a distance of 30.5 kpc \citep{siegel11} and $E(g-r)$=0.11 \citep{schlafly11}, is shown as a solid black line. The GMOS field-of-view is 5.5\arcmin a side. {\bf Right panel:}  \textit{HST}/ACS photometry of NGC 6652 and the discovered variables within its field-of-view (202\arcsec$\times$\,202\arcsec). \label{fig1}}
\end{figure*}

NGC 6652 also lies in front of the Sagittarius dwarf spheroidal  (Sgr dSph, hereafter Sgr) leading tidal arm, some 4 degrees away from the center of Sgr, and therefore deep imaging of the cluster has been frequently used to characterize the Sgr stream \citep{siegel11,sohn15}.

In this paper we use archival Gemini imaging to conduct a search for variable stars in NGC 6652 and the background Sgr stream.  Section \ref{sec:obs} presents observation and data reduction, while Section \ref{sec:search} describes the search. Section \ref{sec:variables} describes the found variables and reassesses the nature of the known variables, and we finish summarizing at Section \ref{sec:summary}.   

\section{Observations and data reduction} \label{sec:obs}

NGC 6652 was observed with the Gemini South telescope, located in Cerro Pach\'on, Chile, on the night of May 2, 2011 under Gemini program GS-2011A-Q-20  (PI: Heinke). Observations were taken with the Gemini Multiobject Spectrograph \citep[GMOS,][]{hook04} in imaging mode (then equipped with E2V detectors), taking alternatively undithered images in SDSS  $g'$ and $r'$ filters (herefater, $g$ and $r$, for simplicity), with exposure times of 75 secs in both bands. Observations were taken continuously for $\sim6$ hours for a grand total of 86 images in each filter. This cadence and time span are particularly useful to detect short period variables such as  SX Phoenicis (SX Phe) or $\delta$ Scuti stars, which have pulsation periods of the order of just a few hours. For other pulsating stars such as RR Lyrae stars, which vary in timescales of around half a day, the cadence will not be enough to obtain a period although the time span is enough to recognize them as variable stars. The GMOS fov is 5.5$\times$ 5.5 arcmin, with a pixel scale of 0.146\arcsec when reading with a 2$\times$2 binning.

These images were originally obtained to study the optical counterpart to the low-mass X-ray binary XB 1832-330 in NGC 6652 \citep{engel12}, but ignoring the rest of the cluster.  

Raw science frames together with their associated calibrations were retrieved from the Gemini Observatory Archive$^1$\footnote{$^1$\url{https://archive.gemini.edu}}. Bias subtraction, flat fielding and trimming of the unilluminated sections of the CCDs were done using the GMOS specific tasks in the IRAF/Gemini package. FWHM of the psf was measured with the gemseeing task. Median FWHM of the complete dataset are 0.68\arcsec and 0.61\arcsec\, for $g$ and $r$, respectively. 

\section{Variable star search and photometry}\label{sec:search}

\begin{figure*}[t]
    \centering
    \begin{minipage}{.48\textwidth}
    \vspace{-0.65cm}
              \includegraphics[width=\linewidth]{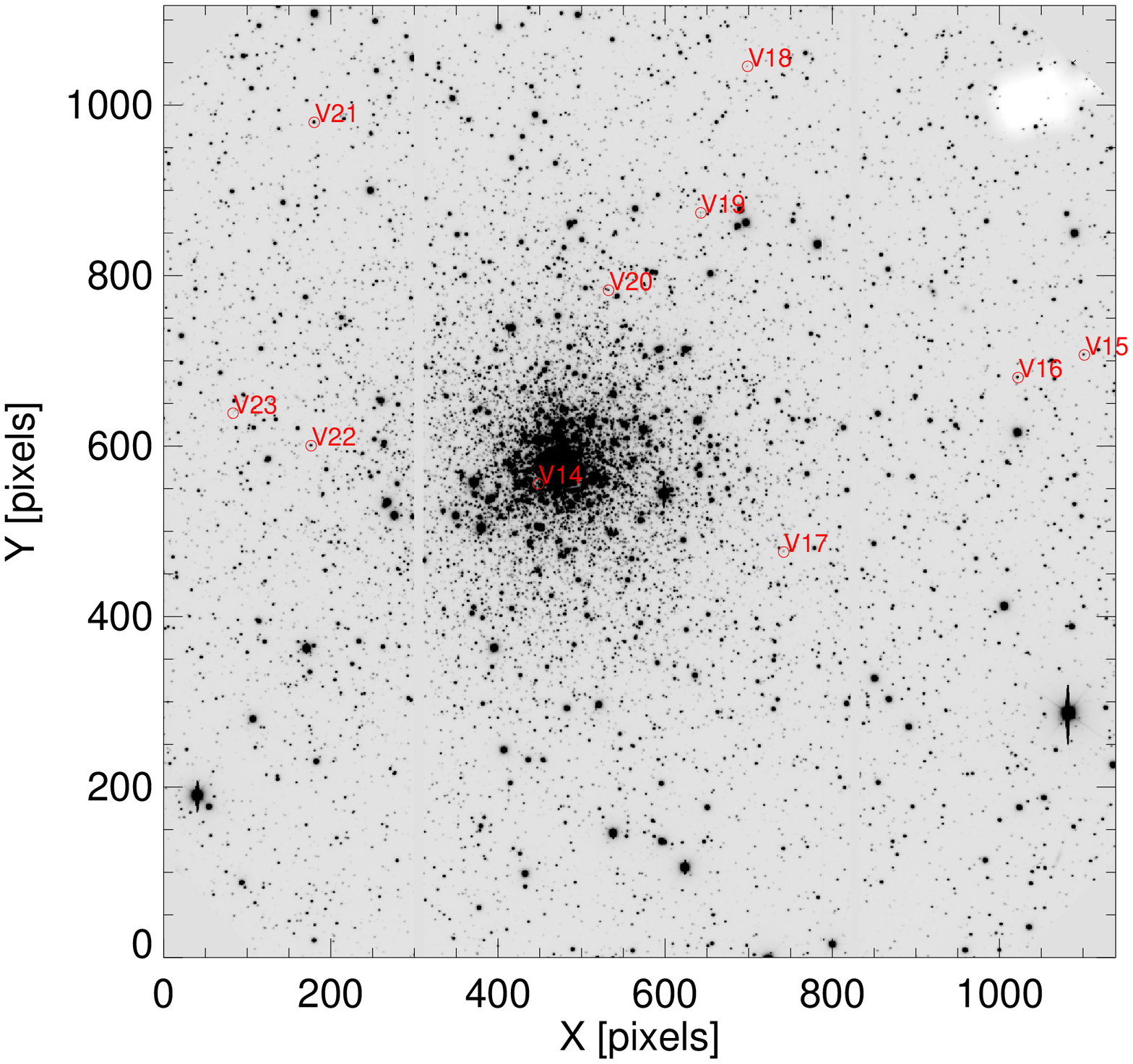}
    \end{minipage}
    \begin{minipage}{0.48\textwidth}
        \includegraphics[width=\linewidth]{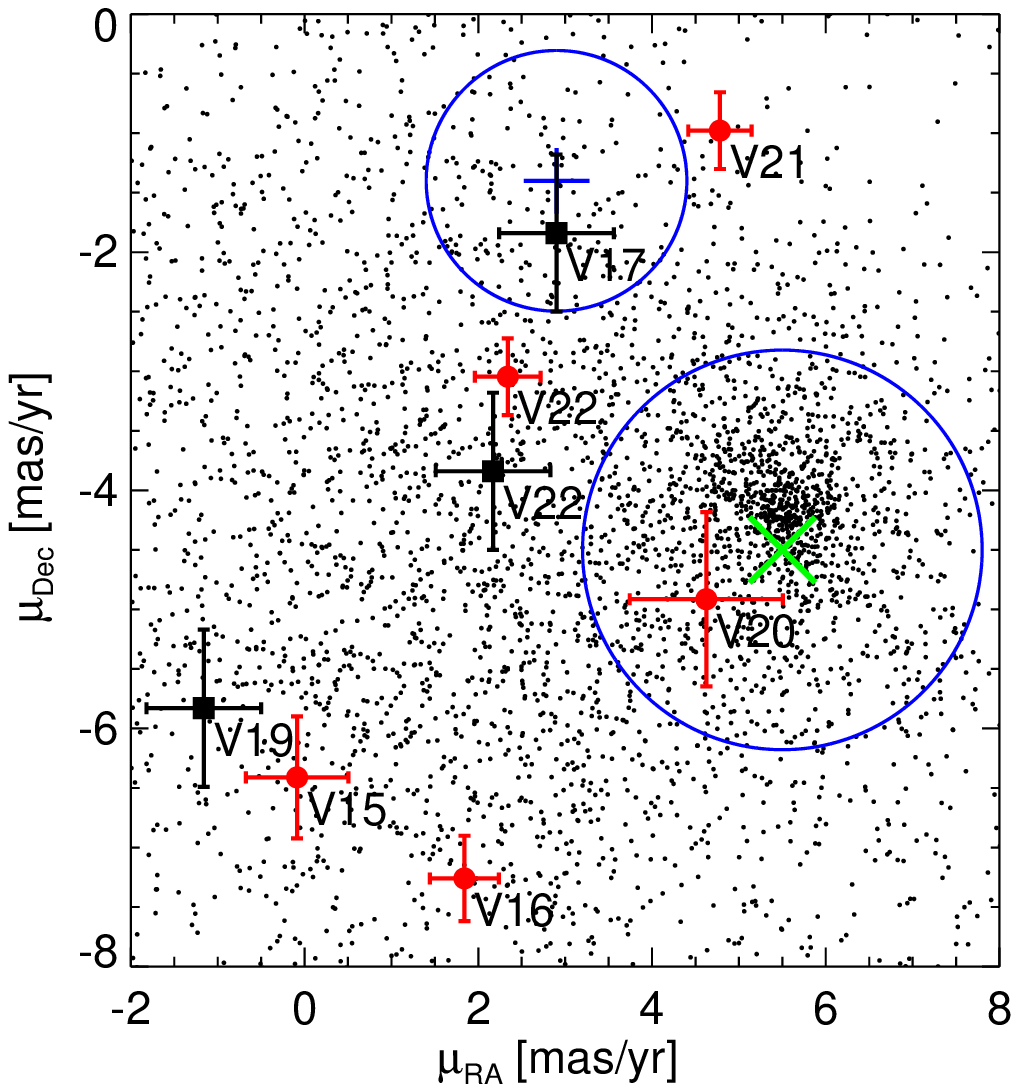}
    \end{minipage}
 \caption{Left panel:  A finding chart for the discovered variables in the NGC 6652 field, based on a GMOS $r$ image. The position of V14 is also indicated. The fov is $5.5\arcmin\times5.5\arcmin$. North is up, East to the left. Right panel: proper motions of stars in the NGC 6652 field based on \textit{Gaia} observations. The green X symbol indicates the proper motion of NGC 6652, while the upper blue circle indicates the proper motion for stars in the background Sgr field, both measured by \citet{sohn15}. The solid red circles indicate the proper motions of the discovered variables present in the \textit{Gaia} DR2, while the black squares indicate the \citet{soto17} \textit{HST} proper motions.  The large blue circle is the criterion of 2.3 mas yr$^{-1}$ used by \citet{soto17} to define cluster membership. For sake of clarity, V14 is not included since its proper motion coincides exactly with that of the cluster.}   
 \label{fig2}
\end{figure*}

Variables stars were searched using the difference imaging program ISIS \citep{alard00}. 
The process involves several steps made separetely for each filter. First, all images are registered to one image that serves as astrometric reference. This image is usually the one with the best image quality in the sample. A small set of the best images (best behaved psf, low background, no artifacts) are then combined to form a photometric reference. ISIS then finds a convolution kernel, that when applied to the reference frame will match the psf, but also compensate for different exposure times, variations in background and airmass, to each individual image of the set. Once the psf is matched, the subtraction is applied. The residual images are then combined obtaining the mean of the absolute normalized deviations (a variance image), where variable residuals (e.g. genuine variable sources) will stand out.

Once this variance image is constructed, it is visually inspected in order to find intrinsic variations and discard blemishes that can produce similar signals (bad pixels, saturated stars, cosmic rays, etc).  Variable stars leave residuals that preserve a stellar-like shape and therefore can easily be separated from the residuals from ill-subtracted saturated stars or hot pixels. We have found that in the extremely crowded regions of globular cluster cores this visual inspection gives better results than applying a detection threshold to the variance image, which necessarily picks up a large amount of false positives. The success of our approach is best exemplified in the case of NGC 2808 where, using the same dataset, \citet{corwin04} did not found any SX Phe stars while \citet{catelan06} found 4 of these faint, low amplitude variables. We applied this method in several publications with excellent results \citep{salinas07,salinas16,salinas18}.  In the case of NGC 6652, our visual inspection revealed the existence of 14 candidate variable sources, which further inspection of their light curves narrowed down to nine (see below).

The final step of ISIS is to conduct psf photometry over the selected candidates. This photometry is given in fluxes relative to the image chosen as reference. In order to transform these relative flux light curves to magnitudes, standard psf photometry has to be performed over the reference images.

Psf-fitting photometry of the reference image was done using DAOPHOT/ALLSTAR \citep{stetson87}, following standard procedures, including the selection of $\sim$50 bright isolated stars per filter to determine the psf.  For the variable sources discovered with ISIS, the relative flux light curves were associated with the ALLSTAR results to convert them to magnitudes using the procedure described in \citet{catelan13}.

Calibration of the magnitudes to the standard system was achieved using the color terms given in the Gemini web pages$^2$\footnote{$^2$\url{http://www.gemini.edu/sciops/instruments/gmos/calibration/photometric-stds}}, but the zeropoints derived by \citet{engel12} from standard stars observed the same night of the observations. The calibrated magnitudes of all stars in the GMOS field, together with the intensity-weighted magnitudes for the new variables can be seen in Fig. \ref{fig1}, left panel. Saturation starts slightly above the horizontal branch, and it is unlikely that affects the search for RR Lyrae stars (see below). Additionally, \textit{HST}/ACS photometry from \cite{sarajedini07} is shown in Fig. \ref{fig1}, right panel, where this time the magnitude of the variables comes from single exposures. \textit{HST} photometry is expected to be mostly unaffected by blends and therefore gives an independent measurement to assess the results from the GMOS photometry. Finally, periodicity in the light curves was searched using the phase dispersion minimization algorithm \citep{stellingwerf78} as implemented in IRAF, and the astrometry of all sources was established cross-correlating with known 2MASS sources via SCAMP \citep{bertin06}.

\section{Known and new variable stars}\label{sec:variables}

\subsection{Proper motion sources}

As an aid to the classification of the discovered variables we used proper motions from two sources, the \textit{HST} proper motions from \citet{soto17} and the recent \textit{Gaia} DR2 \citep{gaia,gaiadr2}. \citet{soto17} released only coarse proper motions based on \textit{HST}/ACS and \textit{HST}/UVIS images of the cluster separated by 7.5 years, with the advantage that observations are deeper than \textit{Gaia} and can also identify stars in the very core of the cluster.  As in \citet{soto17} we adopt the criterion of a proper motion relative to the cluster center of less than 0.35 pixel (2.3 mas) as indicator of cluster membership. Since \citet{soto17} gave only proper motions relative to the cluster center, we used the NGC 6652 proper motion measured by \citet{sohn15} ($\mu_{\rm RA}$=5.5 mas/yr , $\mu_{\rm Dec}$=--4.5 mas/yr) to give a zero point to the \textit{HST} proper motions. For variables outside the \textit{HST} field of view we have complementary used proper motions from the \textit{Gaia} DR2 \citep{gaia,gaiadr2}. 

Fig. \ref{fig2}, right panel, shows \textit{Gaia} proper motions for stars in the GMOS fov (small black dots). The cluster of points around $\mu_{\rm RA}$=5.5 mas/yr and $\mu_{\rm Dec}$=--4.3 mas/yr shows the proper motion of stars in NGC 6652. The proper motion measured by \citet{sohn15} is shown with a green cross, and it has only a small offset with respect to the \textit{Gaia} results. The proper motions for the discovered variable stars can be seen with black  symbols corresponding to \textit{HST} data, and red circles for \textit{Gaia}. The circle centered at $\mu_{\rm RA}$=2.9 mas/yr and $\mu_{\rm Dec}$=--1.4 mas/yr is the proper motion of the background Sgr field measured by \citet{sohn15}.  Note that  a couple of variables (V18 and V23) do not have proper motions because they are outside the \textit{HST} fov and are too faint to be detected by \textit{Gaia}.

\begin{deluxetable*}{lcccrrc}
\tablewidth{0pt}
\tablecaption{Data for known variables, excluding V14, in the NGC 6652 field from the \citet{clement01} catalogue, plus proper motions from \textit{Gaia}. The classification is, CV: cataclismic variable, E: eclipsing binary, L: long-period variable, LMXB: low-mass X-ray binary, RRab/c: RR Lyrae, SR: semi-regular and U: variable with unknown classification. The last column indicates cluster membership based on the \citet{gaiadr2} proper motions. The long dash, ---, indicates the star has not been found in \textit{Gaia}, and therefore its membership is unknown (marked with a   ``?'' sign).\label{table1}}
\tablehead{
\colhead{ID} & \colhead{RA (J2000)} & \colhead{Dec (J2000)}& \colhead{Type} & \colhead{pm$_{\rm RA}^{\rm GAIA}$}& \colhead{pm$_{\rm Dec}^{\rm GAIA}$} & \colhead{Member?}\\
\colhead{} & \colhead{} & \colhead{} & \colhead{} & \colhead{(mas/yr)} &\colhead{(mas/yr)} & \colhead{}
}
\startdata
V1 & 18 35 41.43 & --32 55 47.5 & U &  $3.028\pm0.199$ & $-6.302\pm0.177$ & N\\
V2 & 18 35 26.79 & --32 54 31.3 & L & $-0.612\pm 0.363$ & $-1.261\pm 0.308$ & N\\
V3 & 18 35 19.10 & --32 58 56.8 & RRab & $-0.076\pm 0.141$ &$-5.886\pm 0.127$ & N\\
V4 & 18 35 31.76 & --32 54 54.6 &  U & $8.768\pm 0.238$ & $-5.118\pm 0.212$ & N\\
V5 & 18 35 45.47 & --33 05 00.6 & RRab &  $-2.498\pm 0.087$ &  $-5.819\pm 0.077$ & N\\
V6 & 18 35 27.39 & --32 59 04.1 & RRc & $1.247\pm 0.156$& $1.327\pm 0.130$& N\\
V7 & 18 35 32.99 & --32 57 23.3 & SR &  $2.887\pm 0.112$& $-9.951\pm 0.094$& N\\
V8 & 18 36 11.83 & --33 02 21.2 & E?& $-1.500\pm 0.103$ &$-5.675\pm 0.095$& N \\
V9 & 18 36 19.84 & --32 56 40.6 & RRab&$3.967\pm 0.076$ &$-5.330\pm 0.068$& Y\\
V10 & 18 35 43.65& --32 59 26.8 & LMXB & --- & --- & ?\\
V11 & 18 35 44.57 & --32 59 38.3 & LMXB  & --- & --- & ?\\
V12 & 18 35 45.75 & --32 59 23.5 &CV?& --- & --- & ?\\
V13 & 18 35 45.81 & --32 59 35.9 &U& $3.538\pm0.289$ & $-3.370\pm0.265$ & Y\\
\enddata
\end{deluxetable*}

\setlength{\tabcolsep}{5pt}
\begin{deluxetable*}{lccccllllrrlll}
\tabletypesize{\scriptsize}
\tablewidth{170pt}
\tablecaption{Positions, mean magnitudes, amplitudes, periods and classification for the new variables discovered in the NGC 6652 together with the known variable V14. Uncertain amplitudes are indicated with a colon, while the $>$ indicates lower limits for some periods. The classification is SX: SX Phoenicis, RRL: RR Lyrae, E: eclipsing binary and U: variable with unknown classification. The last column indicates cluster membership based on the \citet{soto17} and \citet{gaiadr2} proper motions. The long dash, ---, indicates the star has not a measured proper motion.  A ``?'' sign indicates  tentative membership and/or classification.\label{table2}}
\tablehead{
\colhead{ID} & \colhead{RA (J2000)} & \colhead{Dec (J2000)}& \colhead{$\langle g\rangle$} & \colhead{$\langle r\rangle$} &  \colhead{$A_g$} & \colhead{$A_r$} &\colhead{$P$} & \colhead{Type} & \colhead{pm$_{\rm RA}^{\rm Gaia}$}& \colhead{pm$_{\rm Dec}^{\rm Gaia}$} &\colhead{pm$_{\rm RA}^{\rm HST}$} &\colhead{pm$_{\rm Dec}^{\rm HST}$} & \colhead{Member}\\
\colhead{} & \colhead{} & \colhead{} & \colhead{mag} & \colhead{mag} &\colhead{mag} &\colhead{mag} &\colhead{d} &\colhead{} & \colhead{mas/yr} &\colhead{mas/yr} & \colhead{mas/yr} & \colhead{mas/yr} & \colhead{}
}
\floattable
\rotate
\startdata
V14 & 18 35 46.32 & --32 59 32.8 & 18.691 & 18.422 & 0.26 & 0.29 & 0.19   & SX & ---& ---& 5.5 & --4.5 & Cluster\\
V15 & 18 35 31.16 & --32 58 48.7 & 19.508 & 18.733 & 0.03: & 0.02: &$>$0.3 & U & $-0.086\pm0.59$ &$-6.411\pm0.51$ &--- & --- & Field \\
V16 & 18 35 33.00 & --32 58 56.5 & 18.845 & 18.269 & 0.01: & 0.01: &$>$0.3 & E? &$1.840\pm0.39$ &$-7.259\pm0.35$& ---&--- & Field\\
V17 & 18 35 39.50 & --32 59 56.3 & 20.415 & 20.202 & 0.12 & 0.08 & 0.039 & SX & --- & --- & 2.90 & --1.84 & Sgr\\
V18 & 18 35 40.50 & --32 57 10.0 & 21.221 & 20.521 & 0.18 & 0.15 & $>$0.3 & E&--- & --- & ---& ---& Cluster?\\
V19 & 18 35 41.80 & --32 58 00.2 & 21.719 & 20.466 & 0.10 & 0.06 & $\sim$0.24 & E & ---&---&--1.16&--5.83   &Field\\
V20 & 18 35 44.36 & --32 58 26.7 & 19.297 & 18.902 & 0.23 & 0.20 & 0.15 & E & $4.625\pm0.88$& $-4.915\pm0.73$ &5.5 &--4.5& Cluster\\
V21 & 18 35 52.53 & --32 57 29.2 & 18.437 & 17.619 & 0.04: & 0.03: & $>$0.3?  &U &$4.781\pm0.36$&$-0.978\pm0.32$&---& --- &Sgr?\\
V22 & 18 35 52.61 & --32 59 19.9 & 18.686 & 18.060 & 0.02: & 0.02: & $>$0.3  & E? &$2.339\pm0.37$&$-3.045\pm0.32$&2.17&--3.84& Sgr/Field?\\
V23 & 18 35 54.78 & --32 59 08.8 & 23.100 & 21.804 & 0.36 & 0.24 & $>$0.24 & E&--- & ---&---&---&Cluster?\\
\enddata
\end{deluxetable*}

\begin{figure*}
\centering
\vspace{-1.5ex}
\includegraphics[width=0.99\textwidth]{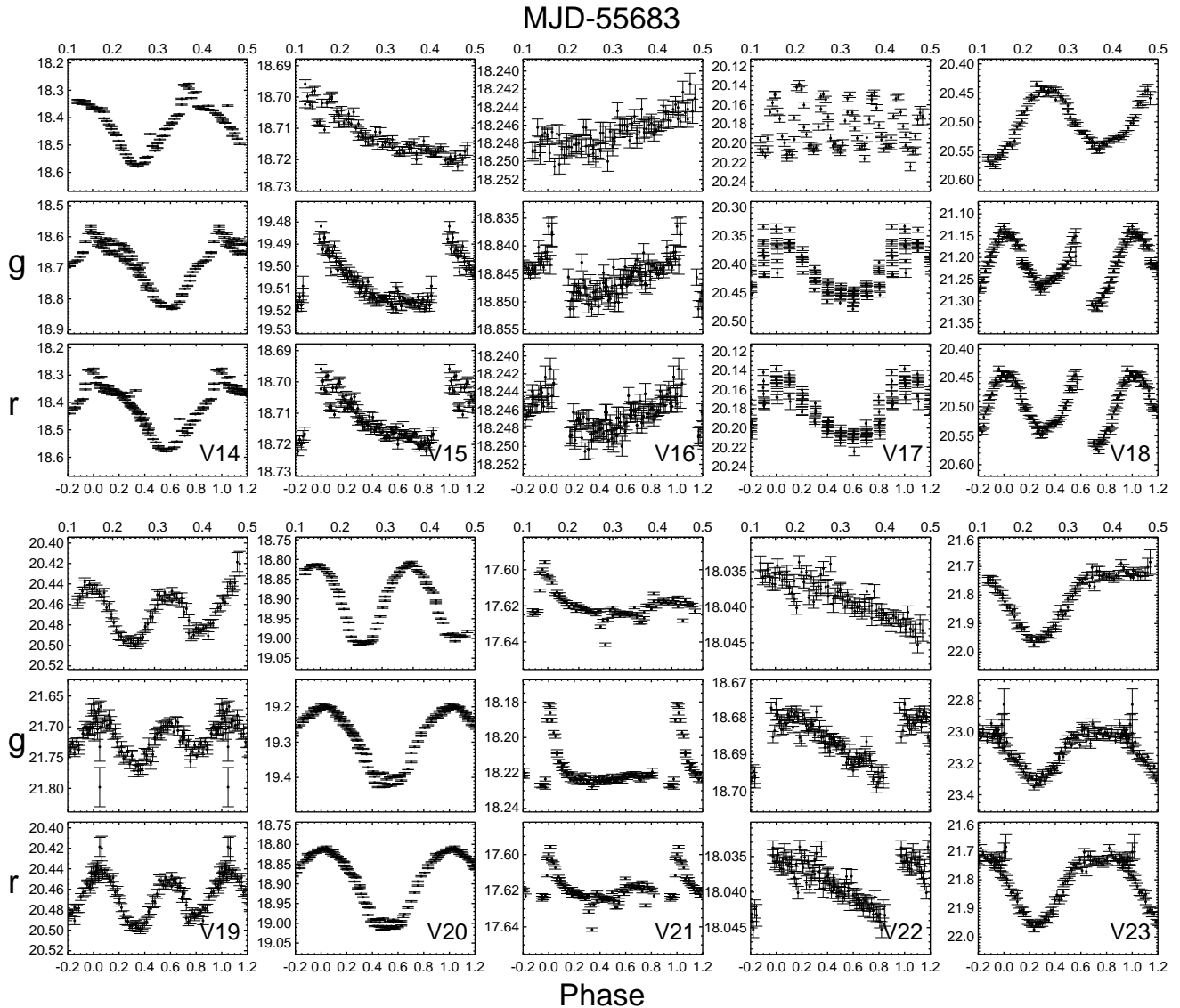}
\caption{Light curves in $g$ and $r$ for the newly discovered variables, plus the known variable V14. Upper panels indicate the $r$ time series photometry in Julian Dates, while the central and lower panels for each panel show the $g$ and $r$ phased light curves, respectively. When only a lower limit period has been found, light curves are given phased to this period.\label{fig3} }
\end{figure*}

\subsection{Known variables outside the GMOS fov}
There have been several searches for variable stars in this cluster \citep{hazen89,heinke01,skottfelt15}. The \citet{clement01} catalogue (updated on January 2015) lists in total 14 variables, of which only V10 to V14  are within the GMOS field of view.

\textit{Gaia} proper motions of all the bright population on the outskirts of the cluster permits us to reassess the membership of the known variables found by \citet{hazen89}. Table \ref{table1} shows the data for V1 to V9 discovered by \citet{hazen89} as given by the \citet{clement01} catalogue. Cross matching between the \textit{Gaia} DR2 sources and the \citet{clement01} catalogue was done with STILTS \citep{taylor06}, using an archival wide-field Blanco/MOSAIC-II image of the cluster (proposal ID 2008A-0289) to check when multiple matches occurred. From the \textit{Gaia} proper motions all but one of these variables (V9) are non-members. V9 is a RRab variable with period of 0.63d.

\subsection{Known variables within the GMOS fov}

 Basic details for these variables are given in Table \ref{table1}.

V10 and V11 are LMXBs discovered by \citet{heinke01}. Even though V10 shows no variability in our data, V11 leaves a positive residual, although is very close to a saturated star. Neither of them are found in the \textit{Gaia} DR2 release.

V12, a cataclismic variable \citep[also discovered by ][]{heinke01} at only $\sim$4\arcsec\, from the cluster center is in a zone of extreme crowding where again our dataset fails to find a variable source. This variable is neither found in \textit{Gaia}.

V13 is an X-ray source for which \citet{skottfelt15} finds almost no variability with the exception of small possible flair. This source, associated to an upper RGB star, appears saturated in our images. Their \textit{Gaia} proper motions are consistent with cluster membership.

V14 was discovered by \citet{skottfelt15} (see details in Table \ref{table2}). They found the light curve of this variable has the typical  sinusoidal shape of a RRc, but its low luminosity would indicate is background star, moreover its period of 0.189845 d is very low for a RRL star, although not unheard of$^3$\footnote{$^3$ The latest version of General Catalogue of Variable Stars \citep{samus17} lists two field RRL with periods even lower than this.}. Alternatively they mention this could be a blue straggler in an eclipsing system. Our observations confirm the period and give some support to its nature as a RRc variable. First, the $g$ light curve shows a hump before maximum light, a common feature in RRL produced by shocks waves in the star's atmosphere \citep[e.g.][]{catelan15}, and second, its color and luminosity are consistent with the horizontal branch of the background Sgr stream (Fig. \ref{fig1}, left panel). A puzzling feature is the sharp increase of luminosity near maximum light seen in both bands. Even though increased scatter at maximum light is well documented for RRcs \citep[e.g.][]{kaluzny00}, the amplitude of $r\sim$0.07 mag of this feature is well above the expectations. A visual inspection of the images near maximum light reveals no issues at the position of this variable. The feature is also not visible in the \citet{skottfelt15} light curve.  The final word for its classification comes from its proper motions.  Not only the \citet{soto17} proper motions  put this variable as a cluster member, but also \citet{simunovic16}, in their measurement of proper motions for blue stragglers in 38 globular clusters, also place it as a cluster member. Therefore we conclude this variable is a member pulsating blue straggler, an SX Phe.



\subsection{The discovered variables}

The positions of the newly discovered variables are shown in Fig. \ref{fig2}, left panel. Data for the newly discovered variables are shown in Table \ref{table2}, while light curves in $g$ and $r$ can be seen in Fig. \ref{fig3}. We follow and extend the naming convention of the \cite{clement01} catalogue. 

{\bf V15}: this variable is located close to the the base of the red giant branch of NGC 6652, but with a color redder by about 0.1 mag. It has a small amplitude ($A_g\sim 0.03$) and a long period that cannot be constrained with the present data. Its membership cannot be established from the \citet{soto17} since it is located outside the \textit{HST} fov, but the \textit{Gaia} proper motions put it far away from the proper motion of the cluster. Therefore, we define this star as a non-member with an unknown classification.

{\bf V16}: also located close to the base of the RGB, but this time with a bluer color. It also has a very low amplitude ($A_g\sim0.01$) and a possible period that cannot be constrained with the present data. This very low amplitude on this timescale would only be consistent with some long-period eclipsing binary. \textit{Gaia} proper motions place it as field star.

{\bf V17}: is a short period variable with a well defined period of 0.039 days. The light curve shape is the typical for SX Phe and Delta Scuti pulsators. In Fig. \ref{fig2} we show an 11 Gyr isochrone from the \citet{dotter08} models with [Fe/H]=--1.3 shifted to the distance (and reddening) of the background Sgr stream. V17 appears in the extension of the background Sgr main sequence occupied by these variables. Even though is too faint to be observed by \textit{Gaia}, the Sgr membership is confirmed by the \citet{soto17} proper motions since it falls almost exactly at the Sgr proper motion calculated by \citet{sohn15} (see Fig. \ref{fig2}, right panel). Since the Sgr tails are known to posses a population older than $\sim$ 9 Gyr \citep[e.g.][]{deboer15} we can safely classify it as an SX Phe (a pulsating blue straggler) instead of a Delta Scuti.

SX Phe stars follow known period-luminosity relations \citep{mcnamara95}. From a compilation of 77 SX Phe in Galactic GCs, \citet{cohen12} derives the PL relation: $M_V=-1.640-3.389\log P \pm 0.11 \pm 0.09$. Using the dust maps of \citet{schlafly11} and the \citet{jester05} transformations between SDSS and Bessell filters, we derive a true distance modulus of 16.84, or a distance of 23.4 $\pm$ 2.1 kpc.  A shortcoming of the \citet{cohen12} PL relation is that in does not consider whether the SX Phe pulsate in their fundamental mode or any harmonic, nor the influence of metallicity. If we take the PL relations from \citet{nemec94}, which take into account both the metallicity and the pulsation mode of the SX Phe, we find that if the measured period corresponds to the fundamental period, then the distance would be 19.5 kpc, while if the period corresponds to the first overtone (we can calculate the fundamental period following \citealt{santolamazza01}), then the distance would be 25.2 kpc. In both cases we assume a [Fe/H]=--1.3, corresponding to the bulk of the Sgr population. These results are significantly shorter than the \citet{siegel11} distance modulus of 17.42 (30.5 kpc) based on MS fitting of the Sgr stream population behind NGC 6652. One possibility is that this SX Phe has a significantly lower metallicity; for [Fe/H]=--2.0, the distance for a first overtone pulsator would be 28 kpc, although there is no evidence that the metallicity of the Sgr tails is much different from the main Sgr body \citep[e.g.][]{carlin18}, the Sgr body reaches metallicities as low as -2.0 \citep{mucciarelli07}.  It is therefore unclear whether this short distance is the result of an unusual broadening of the stream or an uncertainty introduced in the calibration and/or transformation to the Bessell system, or if simply this SX Phe has a metallicity close to most metal poor stars in Sgr.

{\bf V18}: has a clear sinusoidal light curve shape with minima of different depths, which make it most likely a Beta Lyra-type eclipsing binary (or EB). It is located slightly above the NGC 6652 MS which agrees with the position a binary member should have. Unfortunately, there are no \textit{HST} nor \textit{Gaia} proper motions to confirm this. 

{\bf V19}: another variable with sinusoidal shape and somewhat uneven minima, although not as pronounced as in V18. It could be eclipsing of the W Uma type, although a Beta Lyra is not excluded. The \textit{HST} proper motions place it as member of the field.

{\bf V20}: has a sinusoidal  light curve very similar to a RRc, but its position in in the CMD is close to the MSTO. With very similar minima, its classification is most likely a WUma eclipsing binary. Both \textit{Gaia} and \textit{HST} proper motions indicate it is a cluster member.

{\bf V21}: has a unusual light curve with a swift increase of $g\sim0.4$ mag (0.2 in $r$) lasting around 30 minutes, followed by a mostly flat curve. It appears to be a flare or other type of active variable, although its position does not match any of the known X-ray sources. From its \textit{Gaia} proper motions it could be considered as marginally consistent with the Sgr stream. This variable has no \textit{HST}-based proper motions. With the current data we cannot give a definite classification to this variable.

{\bf V22}: presents a small, but clear variation with amplitude of $\sim$0.02 mag in the measured period. If a member of NGC 6652, it could be a RGB star in an eclipsing system. Its \textit{Gaia} proper motion is marginally consistent with Sgr stream, but the \textit{HST} measurements locate it rather as a field member. 

{\bf V23}: the light curve is the typical of an eclipsing variable of the Algol type; an eclipse followed by a flat curve. The secondary eclipse is not visible. Even though it has no proper motions from \textit{HST} nor \textit{Gaia}, its position slightly above the MS is a sign of probable cluster membership.

\section{Summary and conclusions}\label{sec:summary}

We have conducted a new variability study of the metal-rich GC NGC 6652 based on archival Gemini-S/GMOS $gr$ data spanning 6 continuous hours. We discovered 9 new variables. We classify seven of them as eclipsing binaries, one as a SX Phe, while the last remains unclassified. Together with  \textit{Gaia} DR2 and \textit{HST} proper motions, we assign membership to these variables finding that the cluster, the Sgr stream and the field each has 3 of these variables. Using PL relation for the SX Phe found in the Sgr stream we estimate a distance of 23.4$\pm$2.1 kpc to the stream. Finally, we reassess the membership of previously known variables in the cluster thanks to \textit{Gaia} proper motions, finding that only one of the four candidate RRL would belong to the cluster. \textit{Gaia} will undoubtedly clarify the membership of the bright variable star content of the majority of GCs in our Galaxy.

\acknowledgements

We thank the referee for their careful report which helped us clarify several issues.
RS thanks Camila Navarrete for useful information about the Sagittarius stream. This work has made use of data from the European Space Agency (ESA) mission {\it Gaia} (\url{https://www.cosmos.esa.int/gaia}), processed by the {\it Gaia} Data Processing and Analysis Consortium (DPAC, \url{https://www.cosmos.esa.int/web/gaia/dpac/consortium}). Funding for the DPAC has been provided by national institutions, in particular the institutions participating in the {\it Gaia} Multilateral Agreement. This research uses services or data provided by the Science Data Archive at NOAO. NOAO is operated by the Association of Universities for Research in Astronomy (AURA), Inc. under a cooperative agreement with the National Science Foundation.

\facility{Gemini:South (GMOS)}

\software{IRAF \citep{tody86}, ISIS (v 2.1, \citealt{alard00}), stilts \citep{taylor06}, daophot/allstar \citep{stetson87}}

\end{document}